\documentclass[twocolumn]{aa}
\usepackage{graphicx}
\usepackage{natbib}
\bibpunct{(}{)}{;}{a}{}{,}

\newcommand{\kms}{\mbox{$\mathrm{km~s^{-1}}$}}

\newcommand{\Line}[3]{\Ion{#1}{#2}~$\lambda$#3}

\newcommand{\Ion}[2]{#1{\,\scriptsize #2}}

\newcommand{\Ha}{\mbox{${\mathrm H\alpha}$}}
\newcommand{\Hb}{\mbox{${\mathrm H\beta}$}}
\newcommand{\Hg}{\mbox{${\mathrm H\gamma}$}}
\newcommand{\Hd}{\mbox{${\mathrm H\delta}$}}

\begin{document}
\title{Time-resolved photometry and spectroscopy of the new
deeply-eclipsing SW Sextantis star \object{HS\,0728+6738}\thanks{Based
in part on observations obtained at the German-Spanish Astronomical
Center, Calar Alto, operated by the Max-Planck-Institut f\"{u}r
Astronomie, Heidelberg, jointly with the Spanish National Commission
for Astronomy, on observations made with the IAC80 telescope, operated
on the island of Tenerife by the Instituto de Astrof\'\i sica de
Canarias (IAC) at the Spanish Observatorio del Teide, on observations made at the Wendelstein Observatory, operated by the Universit\"ats-Sternwarte M\"unchen, and on observations made with the NASA/ESA Hubble Space Telescope, obtained at the Space Telescope Science Institute, which is operated by the
Association of Universities for Research in Astronomy, Inc., under
NASA contract NAS 5-26555.}}

\author{P. Rodr\'\i guez-Gil\inst{1} \and
        B.T. G\"ansicke\inst{1} \and
	H. Barwig\inst{2} \and
        H.-J. Hagen\inst{3} \and
	D. Engels\inst{3}}

\offprints{P. Rodr\'\i guez-Gil,\\\email{Pablo.Rodriguez-Gil@warwick.ac.uk}}

\institute{
  Department of Physics, University of Warwick, Coventry CV4 7AL, UK
\and
  Universit\"ats-Sternwarte, Scheinerstr. 1, 81679 M\"unchen, Germany
\and
  Hamburger Sternwarte, Universit\"at Hamburg, Gojenbergsweg 112,
  21029 Hamburg, Germany}

\date{Received 2004; accepted 2004}

\abstract{We present time-resolved optical spectroscopy and
photometry, and far-ultraviolet spectroscopy of
\object{HS\,0728+6738}, a cataclysmic variable discovered in the Hamburg
Quasar Survey. We show that the system is a new eclipsing member of
the \object{SW Sex} class of CVs with an orbital period of 3.21
hours. We derive an orbital inclination of $\sim 85 \pm 4\degr$ from the
average eclipse profile, making \object{HS\,0728+6738} the highest inclination
\object{SW Sex} star known.  The optical and far-ultraviolet emission
lines are not or only weakly occulted during the eclipse, indicating
the presence of line-emission sites either far outside the Roche lobe
of the primary or, more likely, above the orbital plane of the binary.
The photometric light curves exhibit fast variability with a period of
$\sim 7$ min, which might be related to the spin of the white dwarf.

 }

\titlerunning{The new \object{SW Sex} star \object{HS\,0728+6738}}
\authorrunning{P. Rodr\'\i guez-Gil et al.}

\maketitle

\section{Introduction}
The SW Sextantis stars compose a group of nova-like (NL) cataclysmic variables (CVs). The NLs lack the characteristic eruptive behaviour shown by their dwarf novae cousins (see \citealt{warner95-1} for a general review on NLs and CVs in general). The \object{SW Sex} stars display a number of so far poorly understood characteristics, most noticeably a very complex behaviour of their emission lines
\citep{thorstensenetal91-1}. More recently,
\citeauthor{rodriguez-giletal01-1} (\citeyear{rodriguez-giletal01-1},
see also \citealt{rodriguez-gil03-1}) and \citet{hameury+lasota02-1}
independently suggested that the \object{SW Sex} stars may contain magnetic white dwarfs. \cite{rodriguez-giletal01-1} also pointed out the possibility of these systems being intermediate polar CVs (IPs; CVs with magnetically truncated discs and asynchronously rotating primaries) with the highest mass transfer rates.

The combination of the exotic but consistent behaviour of the
known \object{SW Sex} stars and their strong clustering in the $3-4$~hour
orbital period range may very well be an important clue for a global
understanding of CV evolution. Nevertheless, some NLs above the $3-4$~hour period interval are found to exhibit some of the defining features of the \object{SW Sex} stars, like \object{BT Mon} \citep{smithetal98-1} and, more recently, \object{RW Tri} \citep{grootetal04-1}. Beyond the realm of CVs, some Low Mass X-ray Binaries (LMXBs) may show distinctive \object{SW Sex} behaviour, such as e.g. \object{XTE\,J2123--058} \citep{hynesetal01-1}.

Despite being intrinsically luminous, most \object{SW Sex} stars are rather
inconspicuous, displaying no outbursts and being weak X-ray
emitters. In fact, about one third of the currently known \object{SW Sex}
stars have been discovered as fairly bright blue objects in the
Palomar-Green survey \citep{greenetal86-1}.

We are currently carrying out a large-scale search for CVs based on
the spectroscopic hallmark of most of them: the presence of strong
emission lines in their spectra \citep{gaensickeetal02-2}. Among the
CVs followed up in more detail \citep[e.g.][]{gaensickeetal00-2,
nogamietal00-1, araujo-betancoretal03-2, rodriguez-giletal04-1,
gaensickeetal04-1}, our survey has so far produced two new
SW\,Sextantis stars \citep{szkodyetal01-1,gaensickeetal02-3}.  Here we
report the discovery of a third new and deeply eclipsing member of the
class: \object{HS\,0728+6738}.

\section{Observations and data reduction}

\begin{table}[t]
\caption[]{\label{t-obslog}Log of Observations}
\setlength{\tabcolsep}{1.1ex}
\begin{flushleft}
\begin{tabular}{lcccc}
\hline\noalign{\smallskip}
Date & Coverage & Filter/Grating & Exp. & Frames \\   
 & (h) & & (s) & \\   
\hline\noalign{\smallskip}
\multicolumn{5}{l}{\textbf{Calar Alto 3.5\,m, TWIN Spectroscopy}} \\
2002 Dec 02    & 1.73   & T05/T06 & 600 & 10 \\
2002 Dec 03    & 8.51   & T05/T06 & 600 & 23 \\
2002 Dec 04    & 1.87   & T05/T06 & 600 & 12 \\
\hline\noalign{\smallskip}
\multicolumn{5}{l}{\textbf{Hubble Space Telescope 2.4\,m, STIS spectroscopy}} \\
2003 Mar 19    &        & G140L  & 730 & 1 \\
\noalign{\smallskip}
\multicolumn{5}{l}{\textbf{Wendelstein 0.8\,m, CCD Photometry}} \\
2001 Apr 01    & 4.19 &  $B$  & 240&  56  \\
2001 Apr 23    & 2.79 &  $B$  & 240&  38  \\
2001 Apr 24    & 4.63 &  $B$  & 60 & 109  \\
2001 Apr 26    & 1.56 &  $B$  & 120&  39  \\
2001 Apr 29    & 1.16 &  $B$  & 60 &  40  \\
2001 May 01    & 1.44 &  $B$  & 60 &  55  \\
2001 May 02    & 4.08 &  $B$  & 60 & 141  \\
2001 May 04    & 4.06 &  $B$  & 60 & 161  \\
2001 May 09    & 1.32 &  $B$  & 60 &  47  \\
2001 May 26    & 1.03 &  $B$  & 60 &  31  \\
2001 Jun 07    & 1.17 &  $B$  & 60 &  63  \\
\noalign{\smallskip}
\multicolumn{5}{l}{\textbf{IAC80 0.8\,m, CCD Photometry}} \\
2003 Sep 24    & 1.35 &  White light  & 20 & 156 \\
2003 Sep 25    & 2.64 &  White light  & 20 & 330 \\
2003 Sep 27    & 1.49 &  White light  & 20 & 182 \\
\noalign{\smallskip}\hline
\end{tabular}
\end{flushleft}
\end{table}

\subsection{Photometry}
\begin{figure}
\centerline{\includegraphics[width=7cm]{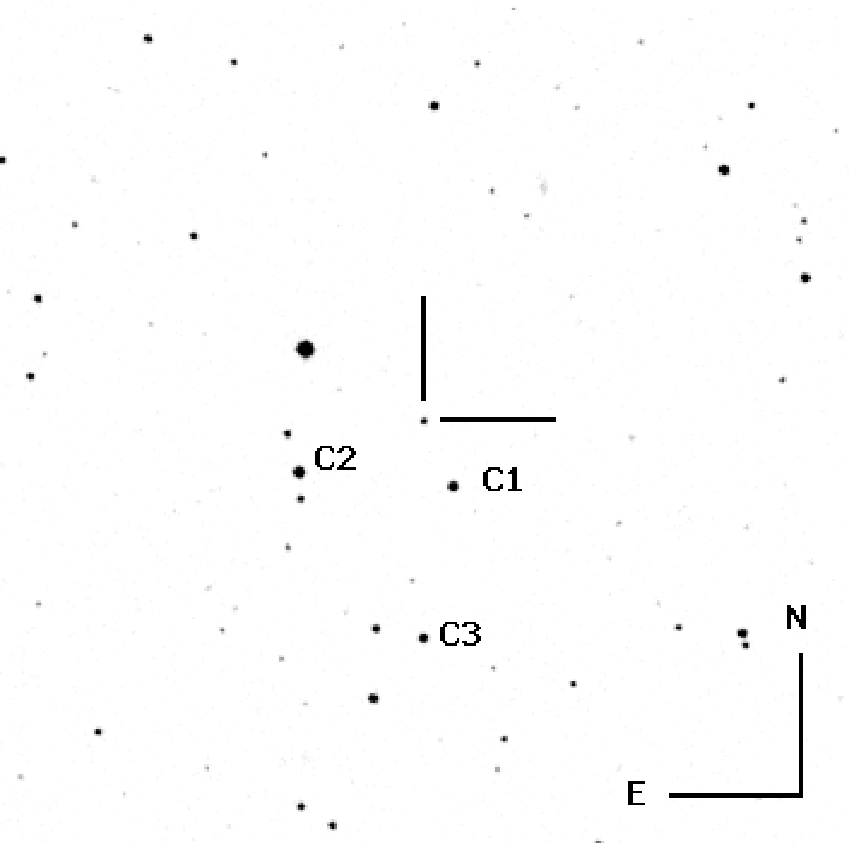}}
\caption[]{\label{f-fc} $10\arcmin\times10\arcmin$ finding chart of
\object{HS\,0728+6738} obtained from the Digitized Sky Survey~2. The
coordinates of the CV are $\alpha(\mathrm{J}2000)=07^\mathrm{h}33^\mathrm{m}41.4^\mathrm{s}$,
$\delta(\mathrm{J}2000)=+67\degr32\arcmin16.2\arcsec$. C1 has been used as
primary comparison star for both the Wendelstein $B$-band and IAC80
white-light CCD photometry.}
\end{figure}

Differential CCD photometry of \object{HS\,0728+6738} was obtained
during 11 nights in the period April to June 2001 with the 0.8-m
telescope at Wendelstein Observatory, Germany, and during 3 nights on
2003 September 24, 25 and 27 with the 0.82-m IAC80 telescope at the
Observatorio del Teide on Tenerife, Spain (Table\,1). At Wendelstein
we used the MONICA CCD camera \citep{roth90-1} which is equipped with
a $1024\times1024$ pixel$^2$ TeK CCD detector. All the images were
taken through a broadband Bessel $B$ filter.
Standard bias subtraction and flat-field correction were done in {\sc
midas}, and aperture photometry was performed using the {\sc
sextractor} \citep{bertin+arnouts96-1}. A more detailed account of the
employed photometry reduction pipeline is given by
\cite{gaensickeetal04-1}. The instrumental magnitudes of
\object{HS\,0728+6738} were derived relative to the comparison star
`C1' (Fig.\,\ref{f-fc}), whilst `C2' and `C3' were used as secondary
comparison stars during different nights.\par

The white-light observations of \object{HS\,0728+6738} at the IAC80
telescope were performed with the Thomson 1024$\times$1024 pixel$^2$
CCD camera using an exposure time of 20 s. We only read out a small
window on the CCD to improve the time resolution which was always
better than 30 s. The individual images were bias-corrected and then
flat-fielded in the usual way. All the data reduction was done with
the {\sc iraf}\footnote{{\sc iraf} is distributed by the National
Optical Astronomy Observatories, which is operated by the Association
of Universities for Research in Astronomy, Inc., under contract with
the National Science Foundation.} package.  The instrumental
magnitudes of the target and three comparison stars were then
calculated using Point Spread Function (PSF) photometry, and
differential light curves were constructed relative to comparison star 'C1' (Fig.\,\ref{f-fc}). Full details of the observations
are given in Table~\ref{t-obslog}.  \par

\subsection{Optical spectroscopy}
Time-resolved spectroscopy of \object{HS\,0728+6738} was performed on
2002 December 2, 3 and 4 with the 3.5-m telescope at Calar Alto
Observatory (Almer\'\i a, Spain). The standard SITe $2000 \times 800$
pixel$^2$ detectors of the TWIN spectrograph were used to
simultaneously image blue and red spectra (gratings T05 and T06,
respectively) at a resolution of 1.2\,\AA~(FWHM; 1.5\arcsec~slit
width). The sampled wavelength ranges were $\lambda\lambda3810-4940$
and $\lambda\lambda6440-7510$ in the blue and red arms,
respectively. A comparison spectrum of a He--Ar lamp was acquired
every three target exposures to guarantee an accurate wavelength
calibration. A log of the spectroscopic observations can be found in
Table~\ref{t-obslog}.\par

The raw images were corrected for the effects of bias and flat-field
structure and then sky-subtracted. The target spectra were optimally
extracted using the algorithm of \cite{horne86-1}. For wavelength
calibration a low-order polynomial was fitted to the arc data, the
{\sl rms} being always $\la 0.05$ \AA. The pixel-wavelength relation
for each target spectrum was obtained by interpolating between the two
nearest arc spectra. These reduction tasks were done with the standard
packages for long-slit spectra within {\sc iraf}. Prior to further
analysis the spectra were normalised by using a low-order polynomial
fit to the continuum.

\subsection{Far-ultraviolet spectroscopy\label{s-hstobs}}
A brief \textit{Hubble Space Telescope}/Space Telescope Imaging
Spectrograph (\textit{HST}/STIS) snapshot observation of
\object{HS\,0728+6738} was obtained on 2003 March 19 as part of a
far-ultraviolet (FUV) spectroscopic survey of CVs
\citep{gaensickeetal03-1}. We used the G140L grating and the
$52\arcsec\times0.2\arcsec$ aperture, covering the wavelength range
$\lambda\lambda1150-1710$ at a spectral resolution of $R\approx1000$.
The data were reduced with the {\sc calstis} (V2.13b) pipeline.\par

\section{The light curve of \object{HS\,0728+6738}}

%\begin{figure}
%\centering
%  \includegraphics[width=9cm]{fig_firstcurve.ps}
%\caption{\label{f-firstcurve} The first light curve of 
%  \object{HS\,0728+6738} contained two consecutive eclipses separated by
%  $\sim 3.2$\,h.} 
%\end{figure}

\subsection{The orbital period}
The first light curve of \object{HS\,0728+6738} immediately revealed
the presence of deep eclipses during which the brightness of the CV
drops by $\sim 2.5$ mag. The observation of two consecutive eclipses
during the first night provided an estimate of the orbital period of
$\sim3.2$ h. 

A precise long-term ephemeris of \object{HS\,0728+6738} was computed
using a total of 14 mid-eclipse times measured in the Wendelstein
and IAC80 light curves (Table~\ref{t-mideclipse}). The eclipse timings
were measured by fitting parabolas to the bottom part of the
eclipses. We have to stress the fact that a number of Wendelstein
light curves lack data at the very eclipse centre. The errors given in
Table\,\ref{t-mideclipse} are purely of statistical nature, the true
error on the eclipse timings may be larger because of the asymmetry in
the wings of the eclipse profiles. A linear least-square fit to the 14
eclipse timings provides the following ephemeris for
\object{HS\,0728+6738}:

\begin{equation}
%\begin{tabular}{rcrl}
T_0(\mathrm{HJD})=2452001.32730(3) + 0.13361946(1)~E~.
%                   $\pm 0.00003$& & $\pm 0.00000001$& \\
%\end{tabular}
\label{eq1}
\end{equation}

\begin{table}
\caption[]{\label{t-mideclipse}Eclipse timings}
\begin{flushleft}
\begin{tabular}{rrrr}
\hline\noalign{\smallskip}
\multicolumn{1}{c}{Time of mid-eclipse} & \multicolumn{1}{c}{Cycle} &  \multicolumn{1}{c}{$O-C$} & \multicolumn{1}{c}{$\sigma(O-C)$}  \\   
\multicolumn{1}{c}{(HJD - 2452000)}&    \multicolumn{1}{c}{E}       &  \multicolumn{1}{c}{(s)} &  \multicolumn{1}{c}{(s)}\\  
\hline\noalign{\smallskip}
\noalign{\smallskip}
1.3275 $\pm$ 0.0003	& 0    & 17.3 & 26.0\\
1.4612 $\pm$ 0.0003	& 1    & 24.2 & 26.0\\
23.5081 $\pm$ 0.0001	& 166  & -2.6 & 9.0 \\
24.4436 $\pm$ 0.0003	& 173  & 11.5 & 26.0 \\
26.3146 $\pm$ 0.0001	& 187  & 39.8 & 9.0 \\
26.4478	$\pm$ 0.0001	& 188  & 3.6  & 9.0 \\
29.3867	$\pm$ 0.0001	& 210  & -59.3& 9.0 \\
32.3269	$\pm$ 0.0001	& 232  & -9.9 & 9.0 \\
32.4609	$\pm$ 0.0001	& 233  & 23.0 & 9.0 \\
34.4650	$\pm$ 0.0001	& 248  & 6.4  & 9.0 \\
39.4087	$\pm$ 0.0001	& 285  & -12.6& 9.0 \\
56.3786	$\pm$ 0.0001    & 412  & 7.1  & 9.0 \\
908.7371 $\pm$ 0.0001	& 6791 & 4.1  & 11.0\\
910.7413 $\pm$ 0.0001	& 6806 & -3.9 & 11.0\\
\noalign{\smallskip}\hline
\end{tabular}
\end{flushleft}
\end{table}

\begin{figure}
\centering
  \includegraphics[width=9cm]{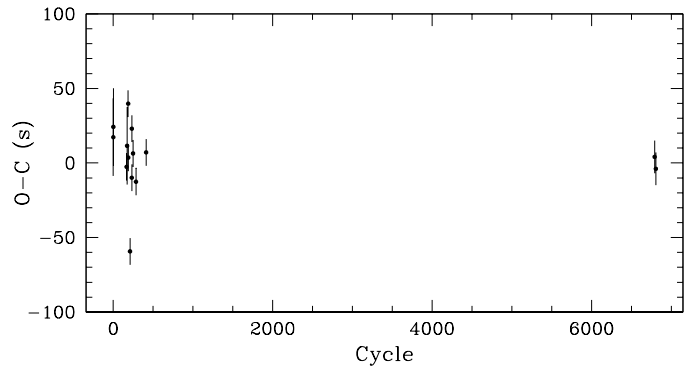}
\caption{\label{f-fig_o_c} $O-C$ diagram of the eclipse timings. The errors have been determined by propagating the errors in the measurement of the mid-eclipse times and the corresponding error from the ephemeris.}
\end{figure}

The observed minus calculated ($O-C$) diagram is shown in
Fig~\ref{f-fig_o_c}. The point with the largest $O-C$ value corresponds
indeed to one of the Wendelstein light curves with poor sampling of
the eclipse centre. Our current data is consistent with a linear
ephemeris.

\begin{figure}
\centering
  \includegraphics[width=9cm]{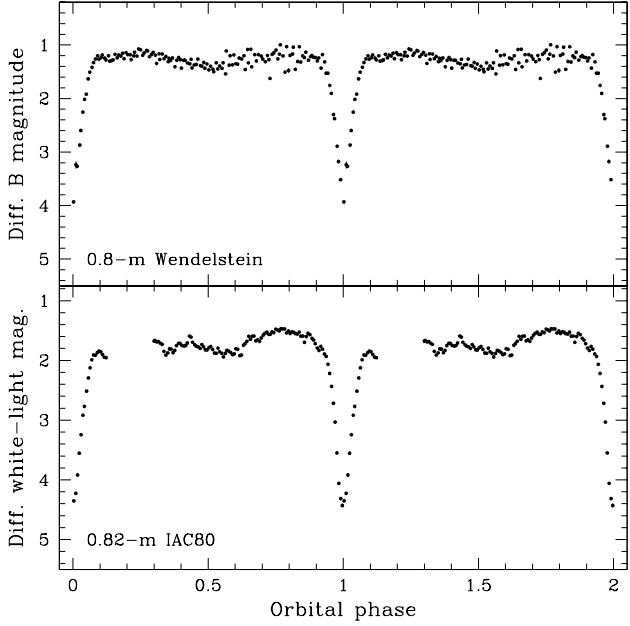}
\caption{\label{fig_avercurves} Average light curves of \object{HS\,0728+6738}. The individual data points were averaged into 150 phase bins. A full orbital cycle has been plotted twice.}
\end{figure}

\subsection{Light curve morphology}

The average Wendelstein ($B$ band) and IAC80 (white light) light
curves are shown in Fig.~\ref{fig_avercurves}. Orbital phases were
calculated according to the ephemeris given in Eq.~(\ref{eq1}) and the
data were averaged into 150 phase bins. The average eclipse depth in
both $B-$band and white light is $\sim 2.7$ mag, probably indicating a
high orbital inclination. The eclipse lasts for $\sim 0.2$ orbital
cycles, or $\sim 40$\,min. 
%Inspection of the two well-sampled eclipses observed at the IAC80 suggests that both eclipse depth and shape vary with time. The out-of-eclipse brightness displays a minimum around orbital phase $\varphi \sim 0.5$ in both $B$-band and white-light.

\begin{figure}
\centering
  \includegraphics[width=8.5cm]{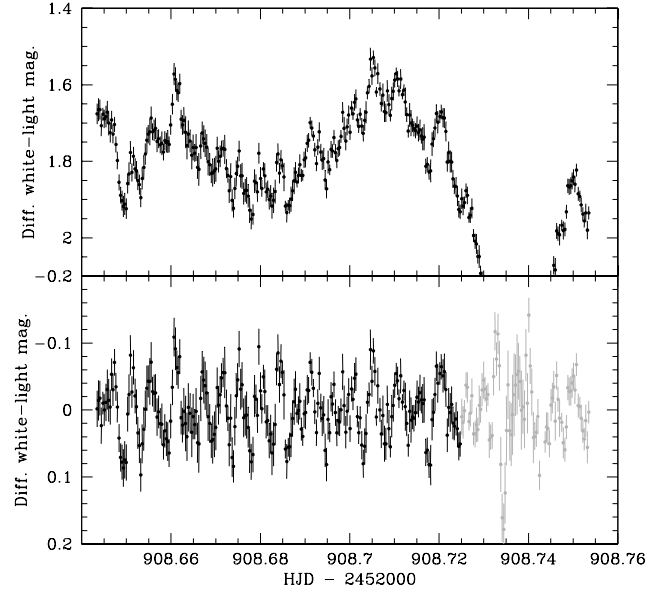}
\caption{\label{fig_qpo} {\em Top panel}: Out-of-eclipse variability on the 2003
September 25 light curve (IAC80 telescope). A short-period oscillation with a time scale of $7-10$ min is easily spotted. {\em Bottom panel}: Same light curve detrended for period analysis. The points plotted in grey lie in the orbital phase interval $0.9-1.1$. See text for details.}
\end{figure}

HS\,0728+6738 shows also significant short-term variability on a time scale of $\simeq7-10$\,min, best resolved in the out-of-eclipse light curve obtained with the IAC80 telescope (Fig.\,\ref{fig_qpo}). Important information on the origin of this variability could be obtained by studying the phase dependence of its amplitude. In order to check if the oscillation is present during eclipse we subtracted a Gaussian fit to the average eclipse profile in white light from the eclipse profile observed on 2003 September 25 (IAC80 telescope). This eclipse-subtracted light curve was then detrended by subtracting the same data set smoothed with a 15-point boxcar. The oscillation is clearly present up to $\varphi \sim 0.96$ and again from $\varphi \sim 0.06$ on. It is difficult to say whether it remains during eclipse or not. The spikes that appear during eclipse could be the residuals of the average eclipse profile subtraction, as the average profile is made up of only two eclipses. A larger set of well-sampled light curves is necessary to address any firm conclusion.
%Unfortunately, the Wendelstein data do not have
%adequate time resolution and sampling to see these variations.

\begin{figure}
\centering
  \includegraphics[width=9cm]{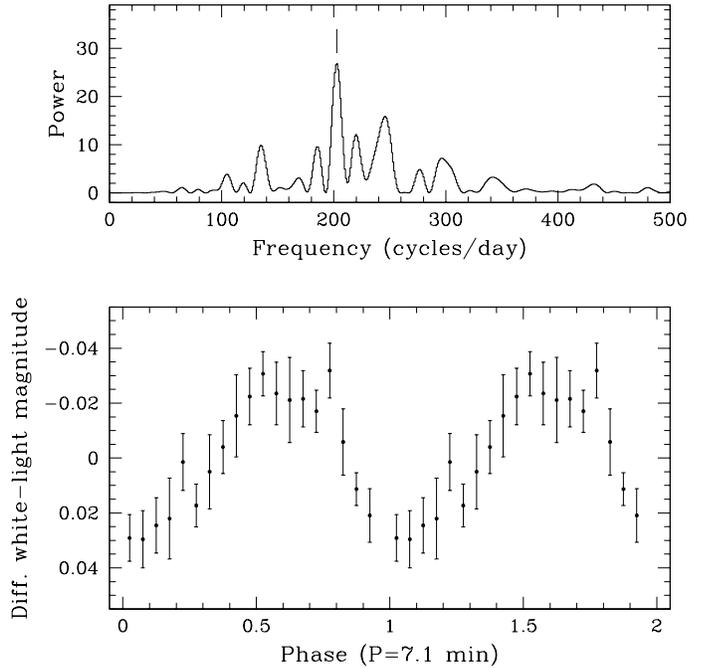}
\caption{\label{fig_qpo_scargle} {\em Top panel}: Scargle periodogram of the
2003 September 25 light curve (IAC80 telescope) after eliminating the eclipse and
detrending. The strongest peak is centred at 7.1 min. {\em Bottom panel}:
the light curve folded on the 7.1-min period after averaging the data
into 20 phase bins (the average value has been subtracted). A complete
cycle has been repeated for continuity.}
\end{figure}

In order to analyse this short time scale variability we decided to eliminate the eclipse
data. A Scargle periodogram \citep{scargle82-1} computed from the detrended
data is shown in the top panel of Fig.~\ref{fig_qpo_scargle}. The
strongest peak is centred at a period of 7.1\,min, confirming the
estimate done by eye. Folding the detrended data on this period
results in a clear quasi-sinusoidal modulation (see bottom panel of
Fig.~\ref{fig_qpo_scargle}). We will discuss the possible origin of
this short-term variation in Sect.\,\ref{sec-shortvar}.

\subsection{Estimate of the orbital inclination}

From the geometry of a point eclipse by a spherical body, it is possible to 
determine the inclination, $i$, of a binary system through the relation

\begin{equation}
\label{eq2}
\left({R_2} \over 
{a}\right)^2=\sin^2(\pi\Delta\varphi_{1/2})+\cos^2(\pi\Delta\varphi_{1/2})\cos^2 i,
\end{equation}

\noindent
where $R_2/a$ is the volume radius of the secondary star, which
depends only on the mass ratio, $q=M_2/M_1$ \citep{eggleton83-1}:

\begin{equation}
\label{eq3}
{{R_2} \over {a}}={{0.49~q^{2/3}} \over {0.6~q^{2/3}+\ln(1+q^{1/3})}}.
\end{equation}

$\Delta\varphi_{1/2}$ is the mean phase full-width of the eclipse at
half the out-of-eclipse intensity. We calculated $\Delta\varphi_{1/2}$
from the $B$-band average light curve (top panel of
Fig.~\ref{fig_avercurves}), assuming an average out-of-eclipse
differential magnitude of $\sim 1.2$. The derived value is
$\Delta\varphi_{1/2} = 0.092 \pm 0.002$.\par

By elimination of $R_2/a$ from Eqs.~(\ref{eq2}) and (\ref{eq3}) we can
obtain an estimate of the orbital inclination of the system after
assuming a value for the mass ratio, $q$. Using as an approximation
the mass-period relation derived by \cite{smith+dhillon98-1}:

\begin{equation}
\label{eq4}
M_2(M_\odot)=0.126~P_\mathrm{orb}({\rm h})-0.11,
\end{equation}

\noindent
where $P_\mathrm{orb}$(h) is the orbital period expressed in hours, we
obtain a value of $M_2 \sim 0.3~M_\odot$ for the secondary in
\object{HS\,0728+6738}. According to \cite{gaensicke97-1} and \cite{smith+dhillon98-1}, the average mass for primaries in CVs above the gap is $M_1 \sim 0.8~M_\odot$. In order to provide an estimate of the uncertainty for the orbital inclination, we will assume that the true primary mass lies in the range $0.6-1.0~M_\odot$. Taking this into account we obtain an orbital inclination of  $i = 85 \pm 4\degr$. Additional systematic uncertainties come from the used $M_2-P_\mathrm{orb}$ relation [Eq.\,(\ref{eq4})] and from the assumption of an axially symmetric accretion disc. The statistical error in the measurement of $\Delta\varphi_{1/2}$ is then negligible. 

As we will show in Sect.\,\ref{s-discussion}, \object{HS\,0728+6738} is
a new \object{SW Sex} star, and we can hence use the $i-\Delta V$
relation of \cite{rodriguez-giletal00-1} to derive an independent
estimate of the orbital inclination. We will assume that the eclipse
depth in the $V$-band does not significantly differ from that in $B$-band.
This assumption is coroborated by the roughly equal depth of the
average Wendelstein $B$-band and the IAC80 white light eclipses. We then get
$i=5.7\,\Delta V+70.6\simeq86\degr$ for $\Delta V \sim 2.7$. This
value agrees well with the estimate derived above from the eclipse
profiles.

%%%
%%% I don't think that the next few lines are necessary...
%%%
%%%For comparison, the \object{SW Sex} star \object{V348 Pup}
%%%has an eclipse depth of $\Delta V \simeq 1.7$
%%%\citep{rodriguez-giletal01-2} and an orbital inclination of $i \simeq
%%%81\degr$ \citep{rolfeetal00-1,rodriguez-giletal01-2}. An inclination
%%%as high as $83\degr$ is not surprising and is consistent with the
%%%large equivalent width of the Balmer lines observed in the average
%%%spectrum of \object{HS\,0728+6738} (see Sect.~\ref{sec-averspectrum}),
%%%the largest seen in a \object{SW Sex} star in normal state.

\subsection{\label{s-longterm}Long-term behaviour}
\begin{figure}
\centering
  \includegraphics[width=9cm]{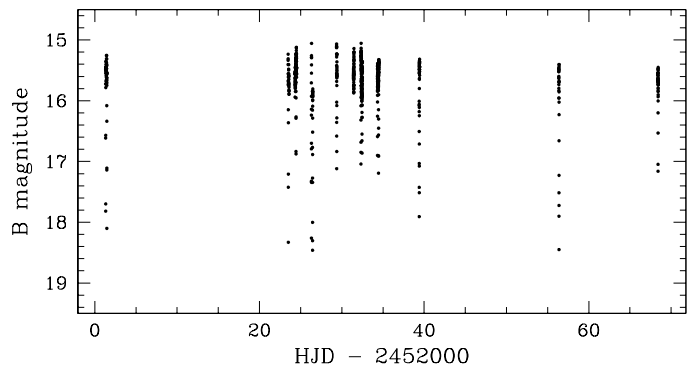}
\caption{\label{f-longterm} Long-term $B$-band light curve of
\object{HS\,0728+6738} from the Wendelstein observations. The
measurements span approximately 67 days. An average value of $B \sim
15.6$ is observed outside of eclipse. The large change in eclipse depth is not real but
due to missing data points at mid-eclipse in many of the light
curves.}
\end{figure}

Fig.~\ref{f-longterm} shows the long-term light curve of
\object{HS\,0728+6738}. The apparent magnitudes of the comparison
stars listed by the USNO-A2.0 catalogue were combined with the
instrumental ones to obtain the (approximate) $B$-band magnitudes of
\object{HS\,0728+6738}. A mean out-of-eclipse magnitude of $B \sim 15.6$ is derived
from the long-term light curve, which agrees with the USNO
measurements of \object{HS\,0728+6738} itself. From our limited data, it is apparent that the out-of-eclipse level changes by $\sim 0.3$ mag on time scales of weeks.

\section{Spectroscopic analysis}

\subsection{The average optical spectrum\label{sec-averspectrum}}
\begin{figure}
\centering
\includegraphics[width=9cm]{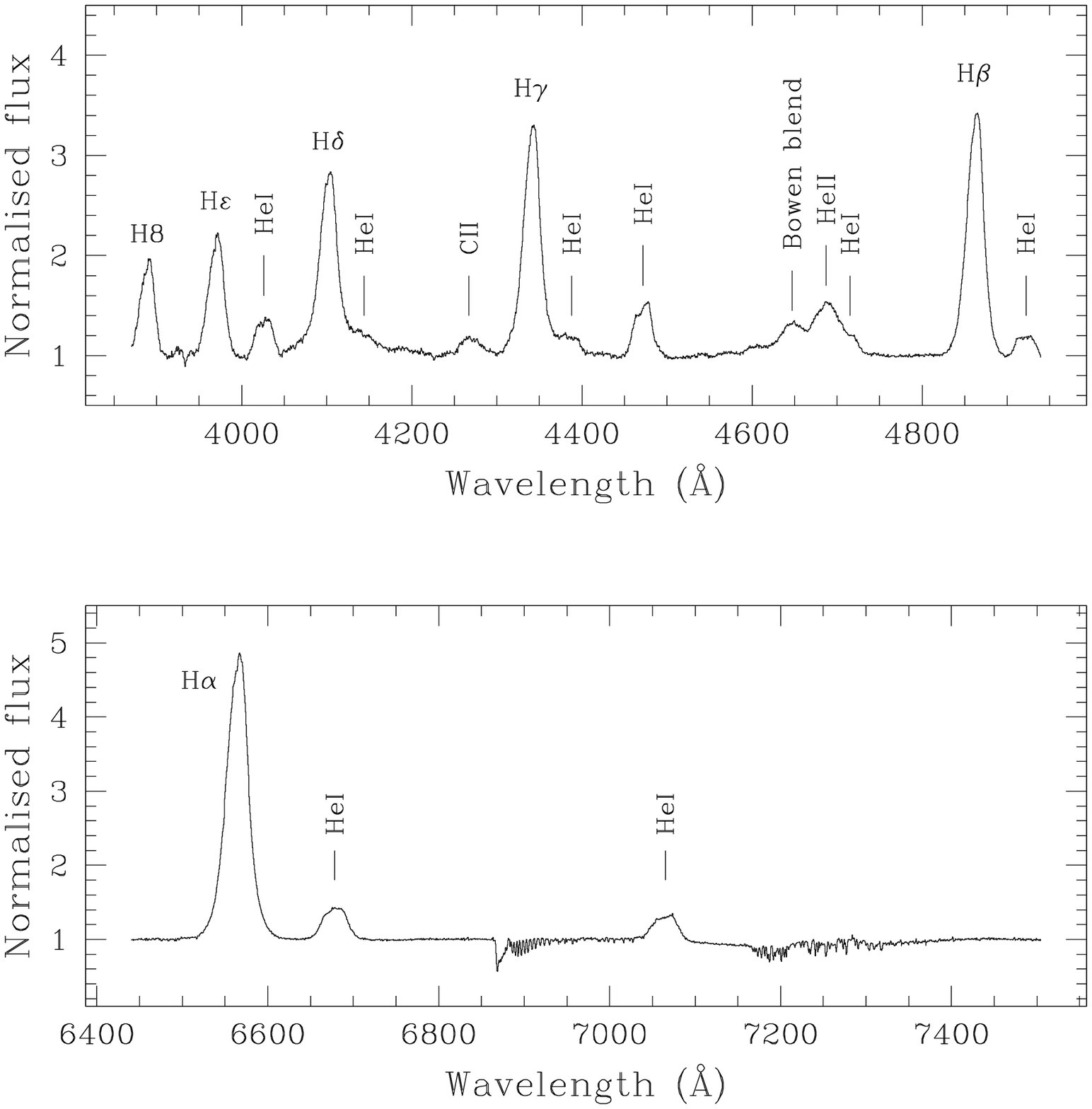}
   \caption{Blue and red average spectra of \object{HS\,0728+6738}.}
    \label{fig_averspec}
\end{figure}

Fig.~\ref{fig_averspec} presents the normalised blue and red average
spectra of \object{HS\,0728+6738}. The emission pattern is dominated
by intense, single-peaked lines of the Balmer series (from H$\alpha$
to H8) and \Ion{He}{I} (like the transitions at $\lambda$4922,
$\lambda$4472 and $\lambda$4026). In constrast to the Balmer lines the
\Ion{He}{I} profiles are almost flat-topped, and double-peaked
profiles start to develop bluewards of \Line{He}{I}{4472}. The
strength of the \Line{He}{II}{4686} and the Bowen blend indicates the
presence of a source of ionising photons. The emission lines of
\object{HS\,0728+6738} also have highly asymmetric profiles with
enhanced wings extending up to $\sim \pm 2000$ \kms~from the line
centre.  Table~\ref{table_lineparam} lists the equivalent widths (EW)
and full-widths at half-maximum (FWHM) of the strongest lines. The
FWHMs were obtained by fitting Gaussians to the line
profiles.  The red spectrum does not contain any spectral feature that
could be ascribed to the photospheric emission of the companion
star. The absorption line bluewards of H$\varepsilon$ is the
\Ion{Ca}{II} K line, which has an interstellar origin and an
equivalent width of $\sim 0.2$ \AA.

\begin{table}[t]
\caption[]{\label{table_lineparam}
Line parameters measured from the average optical spectrum.}
\begin{flushleft}
\begin{tabular}{lrclrc}
\hline\noalign{\smallskip}
Line & EW    & FWHM    & Line & EW    & FWHM   \\
     & (\AA) & (\kms)  &      & (\AA) & (\kms)   \\
\hline\noalign{\smallskip}
\Ha            & $124$ & $1340$ & \Line{He}{II}{4686} & $17$  & 2680     \\
\Hb            & $57$  & 1440   & Bowen blend         & $9$   & 2415     \\
\Hg            & $55$  & 1625   & \Line{C}{II}{4267}  & $5$   & 2090     \\
\Hd            & $49$  & 1770   & \Line{He}{I}{6678}  & $14$  & 1360     \\
H$\varepsilon$ & $27$  & 1570   & \Line{He}{I}{4472}  & $12$  & 1560     \\
H8             & $19$  & 1485   & \Line{He}{I}{4026}  & $8$   & 1730     \\
\noalign{\smallskip}\hline
\end{tabular}
\end{flushleft}
\end{table}

\subsection{Radial velocities \label{sec-rvcs}}
\begin{figure}
\centering
\includegraphics[width=9cm]{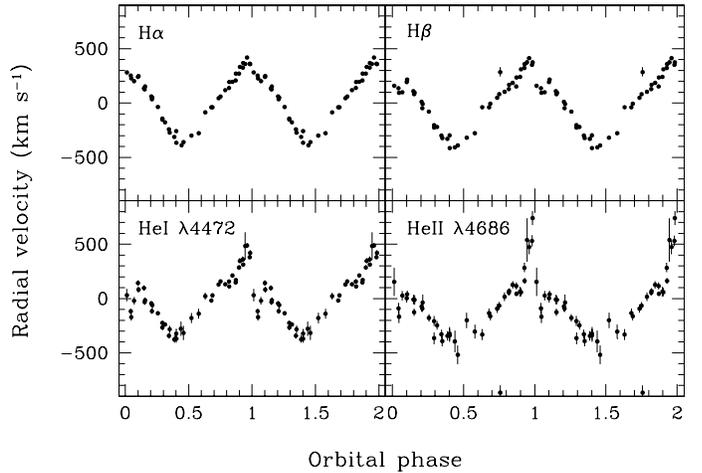}
   \caption{Radial velocity curves of \Ha, \Hb, \Line{He}{I}{4472},
   and \Line{He}{II}{4686} as a function of orbital phase. No
   phase binning has been applied and a full orbital cycle has been
   repeated for continuity.}
    \label{fig-rvcs}
\end{figure}

In order to analyse the effect of the orbital motion on the emission
lines we measured the radial velocities of \Ha, \Hb,
\Line{He}{I}{4472}, and \Line{He}{II}{4686} by convolving the
individual profiles with Gaussian templates. The convolution was
performed on a window suitably selected for not including any
contaminating feature. The FWHM of each Gaussian template was chosen
as the FWHM of the average line profile in each case (see
Table~\ref{table_lineparam}). The radial velocity curves obtained in
this way are shown in Fig.~\ref{fig-rvcs}. We then fitted the velocity
curve of each line with a sinusoidal function of the
form:$$V_r=\gamma-K \sin \left[ 2\pi \left( \varphi-\varphi_0 \right)
\right].$$

\noindent
In Table~\ref{table-rvcfit} we list the resulting fitting
parameters. The absolute values of the $K$ and $\gamma$ velocities should be treated with caution as they depend on the method of the radial velocity measurement. Repeating the analysis described above using a double Gaussian method \citep{schneider+young80-2} with a Gaussian separation of 1000 km s$^{-1}$ yields $K$ velocities larger by a factor of two than those reported in Table~\ref{table-rvcfit}, which agree with the high velocity line wings seen in the trailed spectra diagrams (Fig.~\ref{fig-trailed}). The statistical errors on the velocity measurements are of the order of a few km s$^{-1}$ for both methods. 

All the radial velocity curves are delayed with respect to the assumed
motion of the white dwarf by $\varphi_0 \simeq0.2$. This phase lag indicates that the
main emission site is at an angle of $\sim 72\degr$ to the line of centres between
the centre of mass and the white dwarf. Interestingly, He\,{\sc ii}
$\lambda$4686 is also delayed by approximately the same amount, indicating that the
bulk of this emission is located close in azimuth to the region where Balmer and
\Ion{He}{I} emission comes from.

\begin{table}[t]
\caption[]{\label{table-rvcfit}Radial velocity curve fitting parameters.}
\begin{flushleft}
\begin{tabular}{lccc}
\hline\noalign{\smallskip}
Line & $\gamma$ (\kms) & $K$ (\kms) & $\varphi_0$   \\
\hline\noalign{\smallskip}
\Ha  & $0.8 \pm 0.3$ & $347.6 \pm 0.5$ & $0.2267 \pm 0.0002$  \\
\Hb  & $-23.3 \pm 0.8$ & $344 \pm 1$ & $0.2152 \pm 0.0005$  \\
\Line{He}{II}{4686} & $-21 \pm 4$ & $317 \pm 5$ & $0.189 \pm 0.003$ \\
\Line{He}{I}{4472}  & $-121 \pm 7$ & $255 \pm 11$ & $0.208 \pm 0.006$ \\
\noalign{\smallskip}\hline
\end{tabular}
\end{flushleft}
\rmfamily
\end{table}

All the radial velocity curves displayed in Fig.~\ref{fig-rvcs} show
evidence for rotational disturbance around zero phase, suggesting that
some fraction of the emission line flux originates in an accretion
disc.

%%% ... not necessary ...
%%%
%%%When the eclipse of the disc begins only the disc material
%%%coming towards us is occulted, producing a radial velocity spike to
%%%the red. After mid-eclipse the same is true for the receding disc
%%%material so we get a blue spike in the radial velocities.

\subsection{Equivalent width curves \label{sec-ew}}

\begin{figure}[t]
\centering
\includegraphics[width=9cm]{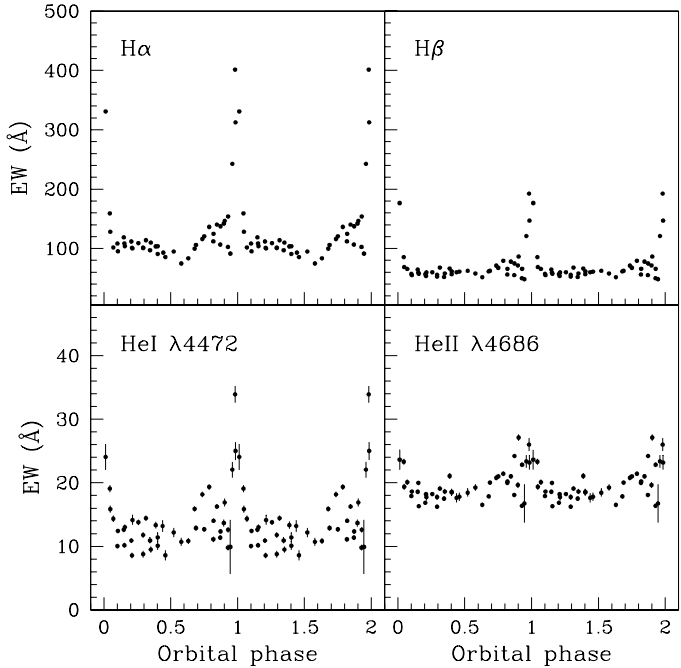}
   \caption{Equivalent width curves of \Ha, \Hb, \Line{He}{I}{4472},
   and \Line{He}{II}{4686} as a function of orbital phase. No phase
   binning has been applied and a full orbital cycle has been repeated
   for continuity.}
    \label{fig-ews}
\end{figure}

\begin{figure}
\centering
\includegraphics[angle=-90,width=8.8cm]{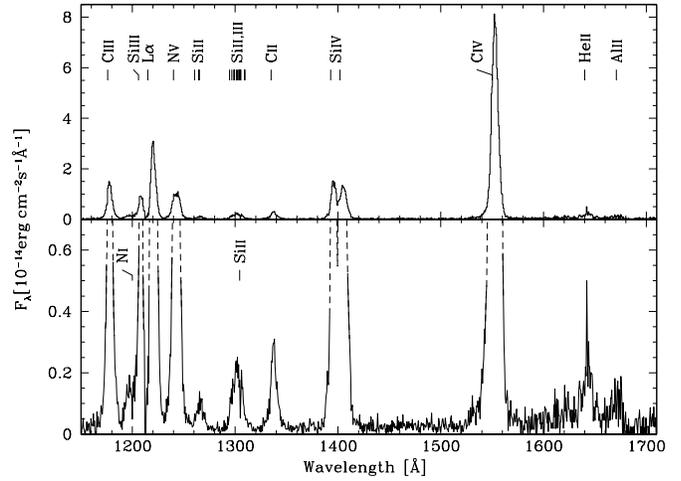}
   \caption{\textit{HST}/STIS far-ultraviolet spectrum of
   \object{HS\,0728+6738} taken at mid-eclipse. Note the absence of
   significant continuum flux and the strong emission lines.}
      \label{f-hst}
\end{figure}

We have computed the equivalent widths (EWs) of the \Ha, \Hb,
\Line{He}{I}{4472}, and \Line{He}{II}{4686} emission lines in each
individual spectrum. In Fig.~\ref{fig-ews} we present the resulting EW
curves as a function of orbital phase. \Ha, \Hb, and
\Line{He}{I}{4472} show strong peaks during the eclipse, indicating
that the regions responsible for the emission of these lines are
eclipsed to a lesser degree than the regions emitting the
continuum. This behaviour requires that the bulk of the Balmer and
\Ion{He}{I} originates either far outside the Roche lobe of the
primary, or situated above the orbital plane. Further support to this
hypothesis is given by the fact that most of the emission lines in the
far-ultraviolet (FUV) remain uneclipsed (see Fig.~\ref{f-hst} and
Sect.\,\ref{s-fuv}). Interestingly, \Line{He}{II}{4686} is eclipsed at
a level comparable to that of the continuum, consistent with the
near-absence of \Line{He}{II}{1640} in the in-eclipse FUV spectrum
(see Fig.~\ref{f-hst} and Sect.\,\ref{s-fuv}).

\subsection{Trailed spectra}

\begin{figure*}[t]
\centering
\includegraphics[width=6.5cm,angle=-90]{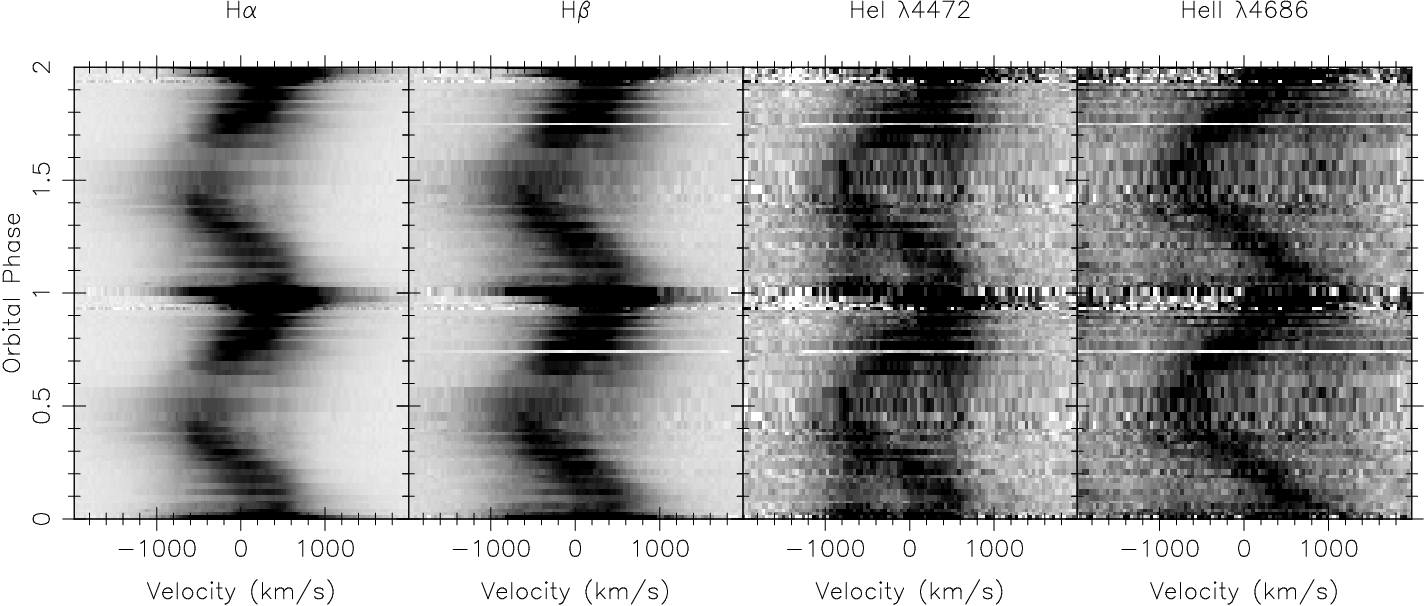}
   \caption{Trailed spectra of \Ha, \Hb, \Line{He}{I}{4472}, and
   \Line{He}{II}{4686}. No phase binning has been applied. Black
   represents emission and a full orbit has been repeated for
   clarity.}
    \label{fig-trailed}
\end{figure*}

Trailed spectra diagrams of the \Ha, \Hb, \Line{He}{I}{4472}, and
\Line{He}{II}{4686} lines were constructed after rebinning the spectra
on to a uniform velocity scale centred on the rest wavelength of each
line. They are presented in Fig.~\ref{fig-trailed}. All the lines are
dominated by a high-velocity emission S-wave which reaches maximum blue
velocity at $\varphi \simeq 0.45$. The Balmer, \Ion{He}{I} and
\Ion{He}{II} emissions reach velocities $\ga 1000$ \kms~at $\varphi
\sim 0$ and $\la -1000$ \kms~half an orbit later. The Balmer and
\Ion{He}{I} emission S-waves are significantly absorbed at $\varphi
\sim 0.5$, but little or no absorption is seen in \Line{He}{II}{4686}.

\subsection{\label{s-fuv}The far-ultraviolet spectrum}
Our brief \textit{HST} observation (Sect.\,\ref{s-hstobs}) has been
obtained exactly during mid-eclipse ($\varphi=0.0$).  The FUV spectrum
of \object{HS\,0728+6738} (Fig.\,\ref{f-hst}) is completely dominated
by emission lines of Ly$\alpha$, C, N, and Si with extremely weak continuum
flux, again indicating that the regions producing both low and high
excitation emission lines are not eclipsed and are located either far
outside the Roche lobe of the primary or above the orbital plane. The
EWs of the strongest lines are given in
Table\,\ref{t-fuv}.  Several narrow lines are detected that are most
probably of interstellar origin (e.g. the \Line{N}{I}{1200} triplet
and \Line{Si}{II}{1304}). Interstellar absorption by neutral hydrogen
is also the likely cause of the dip near 1215\,\AA~between the blue
wing of Ly$\alpha$ and \Line{Si}{II}{1206}. The line flux ratios of the
emission lines are overall typical of normal CVs, with
\Line{C}{IV}{1550} being the strongest line and relatively weak
\Line{N}{V}{1240} emission \citep[see
e.g.][]{maucheetal97-1,gaensickeetal03-1}. Unusual is, however, the
weakness of \Line{He}{II}{1640}, which is consistent with the
fact that the optical \Line{He}{II}{4686} is eclipsed to a similar
degree as the continuum emission (Fig.\,\ref{fig-ews}). Moreover, the
\Line{He}{II}{1640} emission shows evidence for a narrow
component. Our single \textit{HST} spectrum of HS\,0728+6738 suggests
that part of the \Ion{He}{II} emission originates in a region different from
that emitting the bulk of the FUV emission lines. Time-resolved FUV
spectroscopy covering the entire eclipse would be an important probe
to map the various emission regions in HS\,0728+6738.

\begin{table}[t]
\caption[]{\label{t-fuv} Equivalent widths of the emission lines
  contained in the \textit{HST}/STIS spectrum of HS\,0728+6738.}
\begin{flushleft}
\begin{tabular}{lrlr}
\hline\noalign{\smallskip}
Line & EW     & Line & EW    \\
     & (\AA)  &      & (\AA)   \\
\hline\noalign{\smallskip}
Ly$\alpha$                 & 190:$^{1)}$  & \Line{Al}{II}{1670} & $30\pm12$   \\
\Line{He}{II}{1640} & $30\pm12$    & \Line{Si}{II}{1260} & $30\pm8$    \\
\Line{C}{II}{1335}  & $110\pm10$   & \Line{Si}{III}{1206}& 32:$^{1)}$  \\
\Line{C}{III}{1176} & $260\pm20$   & \Line{Si}{III}{1300}& $160\pm15$  \\
\Line{C}{IV}{1550}  & $2000\pm150$ & \Line{Si}{IV}{1400} & $1000\pm80$ \\
\Line{N}{V}{1240}   & $240\pm20$   & \\
\noalign{\smallskip}\hline
\end{tabular}

$^{1)}$ uncertain because of Ly$\alpha$/\Line{Si}{III}{1206} blending and interstellar Ly$\alpha$\ absorption.
\end{flushleft}
\end{table}

\section{\label{s-discussion}HS\,0728+6738 as a new SW\,Sextantis star}

We suggest that \object{HS\,0728+6738} is a genuine SW\,Sextantis star, and we
will compare in the following sections its characteristics to those of
the confirmed members of the \object{SW Sex} class.

\subsection{The optical light curve}
The light curve of \object{HS\,0728+6738} is very similar to those of
the eclipsing \object{SW Sex} stars, such as \object{V348
Pup} \citep{rolfeetal00-1}, \object{DW UMa} \citep{shafteretal88-1} or
\object{V1315 Aql} \citep{dhillonetal91-1}. It exhibits V-shaped
eclipses, short time-scale variations and does not show a significant
pre-eclipse hump due to a conspicuous bright spot (as is the case for the majority of nova-like CVs). 
%The eclipse depth
%of \object{HS\,0728+6738} is variable. Variable eclipse depths are
%also a characteristic of the \object{SW Sex} stars, as e.g. observed in
%\object{DW UMa} (\citealt{biro00-1}; P. Rodr\'\i guez-Gil, unpublished
%data) and \object{PX And} \citep{stanishevetal02-1} and are believed
%to be caused by the precession of an eccentric/tilted accretion
%disc. In fact, a significant number of \object{SW Sex} stars display
%superhumps \citep[see][~and references therein]{pattersonetal02-1},
%photomeric variations with a period a few per cent longer
%(``positive superhumps'') or shorter (``negative superhumps'') than
%the orbital period, which are believed to be one of the hallmarks of
%disc precession/tilting.

The average eclipse depth observed in \object{HS\,0728+6738} is 2.7 mag, which
is the deepest eclipse recorded in any SW\,Sex star (the previous
record holder was \object{V1315 Aql} with an average eclipse depth of
1.9 mag; \citealt{dhillonetal91-1}).

\subsection{\label{sec-shortvar} Short time-scale variability in the light curve}

The fast variability seen in HS\,0728+6738 outside eclipse
resembles the quasi-periodic oscillations (QPOs) on time scales of
$10-30$ min which are detected in many \object{SW Sex} stars
\citep[see][~and references therein]{pattersonetal02-1}.
\cite{rodriguez-giletal01-1} have proposed that the \object{SW Sex}
stars may contain magnetic white dwarfs whose asynchronous rotation
can give rise to observable phenomenology such as short time-scale
variations both in the light curves and the emission lines
(emission-line flaring), and variable circular polarisation. The
accretion scenario they suggested relies on two mechanisms: stream
overflow and magnetic field-stream coupling. By assuming that the
overflown stream couples to the white dwarf's magnetosphere at
approximately the corotation radius one gets a relationship between
the orbital ($P_\mathrm{orb}$) and spin period ($P_1$) such as

\begin{equation}
\label{eq5}
P_1=0.31\, f^{3/2}\, P_\mathrm{orb},
\end{equation}

\noindent
where $f \simeq 0.4-0.6$. Assuming $f\sim 0.5$, we obtain $P_1 \sim
20$\,min for \object{HS\,0728+6738}.  If two-pole accretion is taking
place, we would expect a modulation with a characteristic time scale
of half the spin period, that is, $\sim 10$ min, which is similar to
the time scale of the observed oscillations in \object{HS\,0728+6738}. %Obviously, 
%longer and adquately sampled photometric observations of \object{HS\,0728+6738}
%are necessary to confirm the nature of this short-term
%variability.

\subsection{Spectroscopic behaviour}

The optical spectrum of \object{HS\,0728+6738} is very similar to that
of the deeply-eclipsing \object{SW Sex} stars, e.g. \object{LX Ser}
\citep{youngetal81-2}, \object{BH Lyn} \citep{dhillonetal92-1}, and
\object{V1315 Aql} \citep{dhillon+rutten95-1}. All the \object{SW Sex}
stars show single-peaked emission line profiles. At certain orbital
phases, the lines exhibit central absorption structures which mimic
double-peaked profiles. The strength of these absorption structures
peaks around $\varphi \sim 0.5$ and increases with increasing line excitation level. Also
characteristic of the \object{SW Sex} stars are the highly asymmetric
line profiles. The broad wings are the result of the motion of the
high-velocity S-wave. The velocity semi-amplitude of the S-wave in
\object{HS\,0728+6738} is consistent with that of the S-waves seen in
other eclipsing \object{SW Sex} stars.
%The absorption in the non-eclipsing systems is so strong that the
%\Ion{He}{I} lines are observed completely in absorption around phase
%0.5. Because of the high inclination of \object{HS\,0728+6738} the
%central absorption component is relatively weak in this system.
%%% ... I did not see the point of this paragraph ...
%%%
%%%The high orbital inclination can explain why the S-wave contribution
%%%is so dominant. The picture given by the EW curves and the trailed
%%%spectra together makes sense if we assume that, given the high
%%%inclination of \object{HS\,0728+6738}, the contribution of the disc to
%%%line emission is low. Instead, the lines seem to be almost entirely
%%%dominated by another source of emission which, according to the
%%%behaviour in the optical and the FUV during eclipse, must be located
%%%either far outside of the primary's Roche lobe or above the disc
%%%plane.  

% In non-eclipsing systems like
%\object{LS Peg}, the velocity semi-amplitude can reach $\sim 2000$
%\kms. The increasing amplitude with decreasing orbital inclination
%might indicate line emission from material moving with a significant
%vertical component of the velocity (an idea first suggested by
%\citealt{martinez-paisetal99-1}).

The Balmer, \Ion{He}{I} and \Ion{He}{II} radial velocity curves are
delayed by $\sim 0.2$ orbital cycles with respect to the photometric
ephemeris. This is also a defining feature of the \object{SW Sex} stars.
%But the \Line{He}{II}{4686} radial velocity curve usually
%shares the phasing of the white dwarf motion in these systems. In
%fact, it is widely used to compute an approximate ephemeris for
%non-eclipsing \object{SW Sex} stars such as \object{V795 Her}
%\citep{casaresetal96-1}, \object{LS Peg}
%\citep{martinez-paisetal99-1,tayloretal99-1}, and \object{V442 Oph}
%\citep{hoardetal00-1}. The delay of \Line{He}{II}{4686} suggests a
%common location for the production of the bulk of Balmer, \Ion{He}{I}
%and \Ion{He}{II} emission in \object{HS\,0728+6738}. The
%\Line{He}{II}{4686} radial velocity curve of the eclipsing \object{SW
%Sex} star \object{V348 Pup} also lags $\sim0.2$ orbital cycles behind
%the white dwarf motion. The \Line{He}{II}{4686} line in \object{V348
%Pup} is unusually strong, sometimes stronger than \Hb. This has been
%attributed to enhanced \Line{He}{II}{4686} emission in magnetic
%structures \citep{rodriguez-giletal01-2}.\par

The EW curves of \object{HS\,0728+6738} reveal another key feature of
the \object{SW Sex} stars: the lines are less eclipsed than the
continuum, suggesting an origin either outside the Roche lobe of the primary
or above the main continuum source, i.e. the disc.
% Unlike other eclipsing \object{SW Sex} stars
%(e.g. \object{BH Lyn}; \citealt{hoard+szkody97-1}) the same applies to
%\Line{He}{II}{4686}. This again suggests that \Line{He}{II}{4686}
%originates close to the location where Balmer and \Ion{He}{I} emission
%do, very likely at a certain height above the disc.\par

The trailed spectra show an emission-line behaviour very similar to
that of the deeply-eclipsing \object{SW Sex} stars \object{BH Lyn}
\citep{hoard+szkody97-1}, \object{BT Mon} \citep{smithetal98-1}, and
\object{V348 Pup} \citep{rodriguez-giletal01-2}, with the exception
that \Line{He}{II}{4686} is not dominated by the high-velocity S-wave
in these three objects.
%, giving further support to the idea of the bulk of
%Balmer, \Ion{He}{I}, and \Ion{He}{II} emission in
%\object{HS\,0728+6738} coming from the same region.\par

%%% this repeats basically the first paragraph of 5.3

%%%A defining characteristic of the \object{SW Sex} stars is the
%%%apparition of absorption troughs in the Balmer and \Ion{He}{I} line
%%%cores at $\varphi \sim 0.5$. Only weak absorption is
%%%seen in \object{HS\,0728+6738}. This is not surprising as the strength
%%%of the so-called ``0.5-absorption" in the \object{SW Sex} stars
%%%increases with decreasing inclination and increasing excitation level
%%%(for comparison see the average spectrum and trailed spectra of the
%%%non-eclipsing \object{SW Sex} star \object{V795 Her} in
%%%\citealt{casaresetal96-1}). The very high inclination of
%%%\object{HS\,0728+6738} ($i \sim 83\degr$) then explains the lack of
%%%significant core absorption seen in the trailed spectra.

\section{Conclusions}

In the course of this paper we have provided sufficient evidence to
classify \object{HS\,0728+6738} as a new eclipsing \object{SW Sex}
star. The observed behaviour matches all the conditions needed to be a
member of this class of CVs, namely:
\begin{enumerate}
\item The optical spectrum of \object{HS\,0728+6738} is dominated by
strong, single-peaked Balmer, \Ion{He}{I} and \Ion{He}{II} emission
lines. Intense, single-peaked lines are also observed in the FUV.
\item The EW curves reveal that the emission lines are less obscured
than the continuum during eclipse. The mid-eclipse FUV spectrum shows
almost no continuum but strong emission lines remain, supporting the
hypothesis of the presence of a line-emitting site above the disc.
% Contrary to what happens in other
%\object{SW Sex} stars \Line{He}{II}{4686} also behaves in the same way
%\textbf{Eh? HeII is mostly eclipsed...?}, suggesting that the bulk of
%line emission forms in structures above the accretion disc or outside
%the Roche lobe of the primary.
\item Line emission is dominated by an intense, high-velocity S-wave
reaching velocities of $\sim \pm 1000$ \kms. The S-wave reaches bluest velocity at $\varphi \sim 0.45$, a well-known characteristic of the emission
S-waves in the rest of the \object{SW Sex} stars.
\item The radial velocities of the Balmer, \Ion{He}{I}, and
\Ion{He}{II} lines show the characteristic $\sim 0.2$-cycle phase
delay with respect to the motion of the primary.
% The delay of the
%\Line{He}{II}{4686} radial velocity curve is not common among the
%\object{SW Sex} stars, suggesting that some of the emission forms
%above the orbital plane, close to the regions where Balmer and
%\Ion{He}{I} lines originate.
\item The Balmer and \Ion{He}{I} S-waves show absorption at
$\varphi \sim 0.5$.
% The core absorption typical of the \object{SW Sex} stars around this orbital phase is not significant, probably due to the high orbital inclination of \object{HS\,0728+6738}.
\item The optical light curve displays variations at a time scale of
$\sim 7$ min. It is now accepted that most of the \object{SW Sex}
stars exhibit QPOs in their light curves (see
\citealt{pattersonetal02-1} and references therein), and
\object{HS\,0728+6738} does not seem to be an exception. 
%Nevertheless,
%this point must be confirmed with a more complete photometric
%monitoring.
\end{enumerate}

\begin{acknowledgements}
We thank the anonymous referee for his/her valuable comments on the original manuscript. We are grateful to Otto B\"arnbanter and Christoph Ries for carrying out the Wendelstein observations. PRG and BTG thank PPARC for support through a PDRA and an AF, respectively. The HQS was supported by the Deutsche Forschungsgemeinschaft through grants Re\,353/11 and Re\,353/22. The use of the MOLLY package developed and maintained by Tom Marsh is acknowledged.  
\end{acknowledgements}

\bibliographystyle{aa}
\bibliography{aamnem99,aabib}
%\bibliography{aamnem99,$HOME/tex/Papers/Bibliography/aabib}

\end{document}